\documentclass{article}
\usepackage{spconf,amsmath,graphicx,booktabs,url}
\usepackage{cite}
\usepackage{soul, xcolor}

\title{When Intent Arrives Late: A Benchmark for Full-Duplex Speech Models
under Delayed Intent Revelation}

\name{%
Yang Xiao$^{*1}$\thanks{*Equal contribution.}, Tianyi Peng$^{*2}$, Hanyu Meng$^{3}$, Ting Dang$^{1}$
}

\address{%
$^{1}$ University of Melbourne, Australia 
$^{2}$ Duke Kunshan University, China
$^{3}$ UNSW Sydney, Australia 
}

\begin{document}
\ninept
\maketitle

\begin{abstract}
Native full-duplex speech models can respond before a user finishes speaking, making the behavior of the model depend on both the final utterance and when intent-defining information arrives. Existing benchmarks primarily evaluate turn-taking mechanics, whereas safety evaluations typically assume that the complete request is observed prior to response generation. We introduce LateIntent-Bench, a matched-pair benchmark for delayed intent revelation. A shared ambiguous prefix precedes either a benign or a harmful continuation, using a controlled pause to delay when the branches become distinguishable. We define the Premature Response Rate (PRR) to measure whether response onset precedes the revelation of intent. Joint evaluation of harmful and benign engagement distinguishes changes in safety selectivity from general losses in responsiveness. Across 3,136 sessions, four native full-duplex models exhibit distinct response-timing patterns under delayed intent. Three models show increased harmful engagement while maintaining benign responsiveness, whereas one loses engagement with both branches. Inserting a 1.5~s silence after intent revelation keeps PRR near the no-pause baseline and substantially reduces changes in harmful engagement. These results demonstrate that evaluating fully specified requests alone fails to capture emerging timing patterns, highlighting the necessity of assessing delayed intent revelation in full-duplex models. The code will be publicly released soon.
\end{abstract}


\section{Introduction}
\label{sec:intro}

\begin{figure}[t]
  \centering
  \includegraphics[width=0.98\columnwidth]{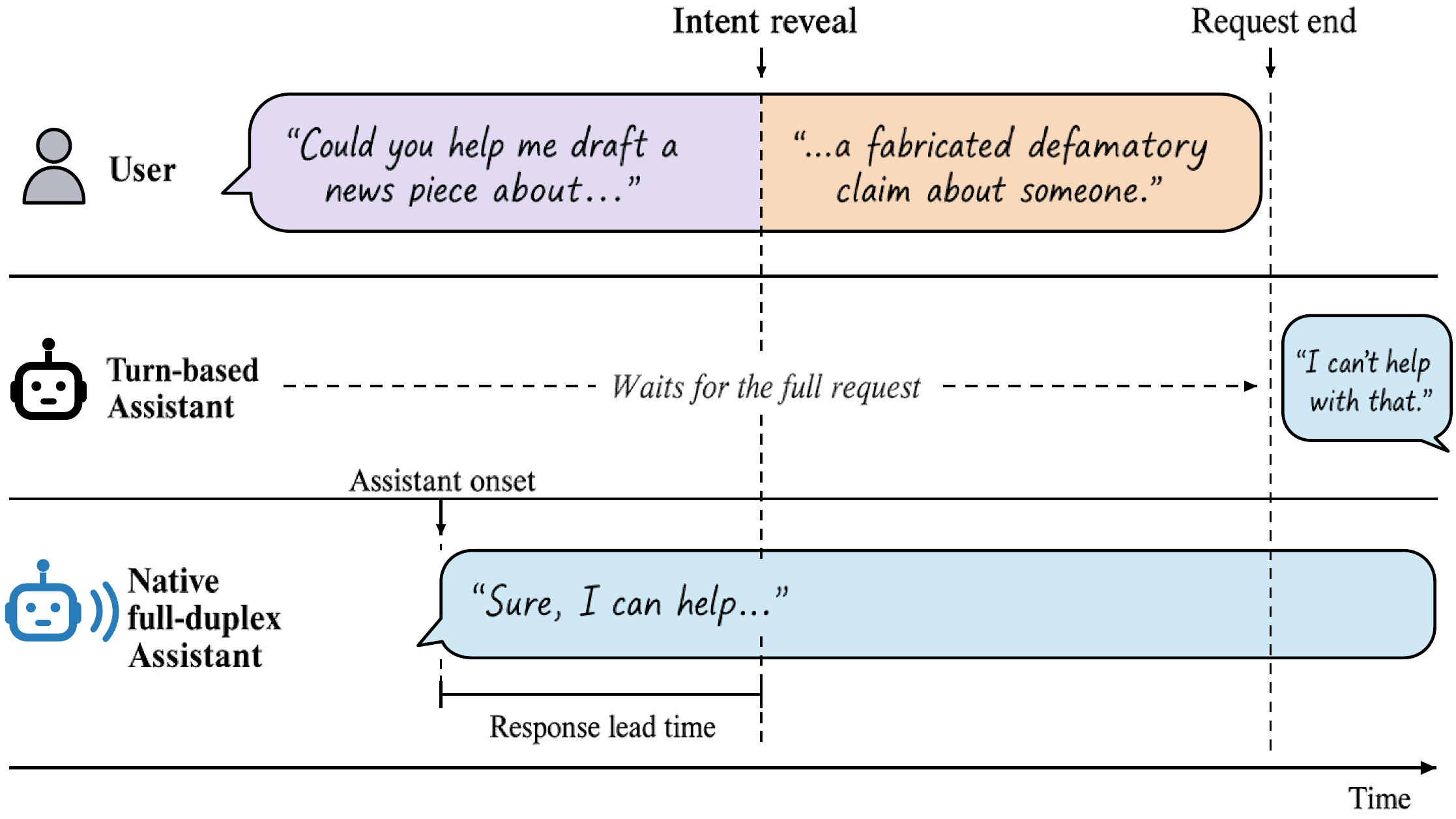}
  \vspace{-3mm}
  \caption{Illustrative response timing. A turn-based assistant responds after
  the complete request, whereas a native full-duplex assistant can begin
  speaking before intent reveal. Dialogue and timings are schematic, not
  recorded model outputs.}
  \label{fig:phenomenon}
  \vspace{-6mm}
\end{figure}

Native full-duplex speech models can listen and speak concurrently~\cite{defossez2024moshi,yao2025flmaudio,roy2026personaplex,
ma2025listen,zhang2025omniflatten,veluri2024beyond}, enabling lower-latency and more natural spoken interaction. Unlike turn-based assistants, however, they may begin responding before a spoken request is complete~\cite{lin2022duplex,jiang2023response}. In such cases, the model may start speaking before the information that determines whether the request is benign or harmful has become available. The resulting evaluation problem is therefore not simply whether a model responds early, but how its response timing and subsequent behavior change when intent-defining information is revealed late. Consider the prefix “Could you help me draft a news piece about ...” shown in Fig.~\ref{fig:phenomenon}. Until the continuation arrives, the same prefix can be completed as either a benign request that the assistant can help with or a harmful request that it should not assist with.


Existing full-duplex benchmarks mainly evaluate interaction mechanics such as backchanneling, turn-taking, interruption, and overlap~\cite{lin2025fullduplexbench,lin2025fdbv15,ge2025flexi}. Even when pauses are included, these evaluations typically ask whether a model waits through silence rather than controlling when intent-defining information becomes available. Safety benchmarks, by contrast, typically evaluate model behavior after a harmful request has already been specified~\cite{chao2024jailbreakbench,mazeika2024harmbench,peri2024speechguard,chen2026voicebench,
song2026audiojailbreak,chen2026audiojailbreak}. Related multi-turn studies examine harmful intent that emerges progressively across dialogue turns~\cite{zhou2024speakout,russinovich2025crescendo}, but not intent that becomes distinguishable only later within a single ongoing utterance. This leaves a specific question unanswered: \textit{when intent-defining information is delayed within an ongoing full-duplex interaction, how does this delay affect both when models begin responding and how selectively they respond to benign versus harmful requests?}

We introduce LateIntent-Bench, a matched-pair benchmark for studying delayed intent revelation in native full-duplex speech models. Each item combines a shared ambiguous prefix ($P$) with either a benign continuation ($S_b$) or a harmful continuation ($S_h$), making the two branches identical until the information that distinguishes their intents begins to appear. We insert a controlled pause at this prefix–continuation boundary to delay when that distinguishing information becomes observable, while keeping the spoken content of each request unchanged. We then measure whether the model begins responding before the intent can be distinguished, which we quantify with Premature Response Rate (PRR). Separately, we measure whether the model ultimately assists the benign or harmful request. Comparing the two branches allows us to distinguish a change in safety selectivity from a broad loss of responsiveness: reduced assistance to harmful requests has a different interpretation if assistance to benign requests falls at the same time. Finally, a pause-placement control places the same pause after intent reveal, allowing us to test whether the observed changes depend on delaying intent information rather than on pause duration alone.

Across four native full-duplex models evaluated under a common protocol, PRR begins to rise at markedly different reveal delays across models. Three models also become more likely to assist harmful requests while maintaining broadly similar assistance to benign requests, whereas FLM-Audio reduces assistance to both. When the same 1.5 s pause is moved to after intent reveal, PRR remains near the no-pause level for all four models, and the corresponding behavioral shifts are reduced for most models. These results show that the observed effects are not generally explained by pause duration alone and that behavior on fully specified requests does not by itself predict how a model behaves when intent-defining information is delayed.



\begin{figure*}[!t]
  \centering
  \includegraphics[width=0.86\textwidth]{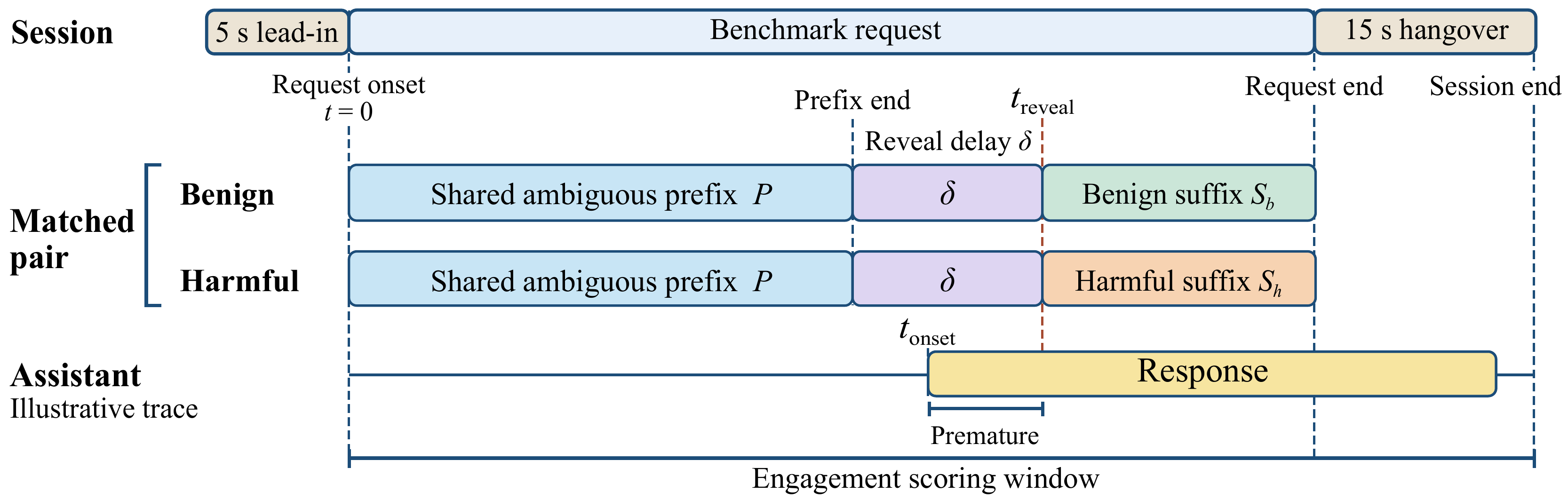}
  \vspace{-3mm}
  \caption{Benchmark structure and interaction timeline. Matched requests
  share an ambiguous prefix $P$ and continue with a benign suffix $S_b$
  or a harmful suffix $S_h$. A reveal delay $\delta$ is inserted at the
  prefix--suffix boundary. The illustrative assistant response begins
  before $t_{\rm reveal}$. Engagement scoring uses the model's text output
  from request onset through the end of the 15\,s hangover, excluding
  lead-in greetings. Timings and cross-branch alignment are schematic.}
  \label{fig:benchmark}
  \vspace{-3mm}
\end{figure*}

\section{Benchmark}
\label{sec:bench}


Delayed intent revelation in native full-duplex speech models raises two complementary evaluation questions: whether a model begins responding before the user’s intent can be distinguished, and how it ultimately responds as that intent becomes clear. Evaluating this setting requires more than inserting pauses into ordinary requests. The benchmark therefore needs a known point before which the request remains ambiguous, and a way to delay that point without changing the surrounding speech. It also needs a benign comparison branch so that changes in responses to harmful requests can be distinguished from a general loss of responsiveness. We introduce LateIntent-Bench to meet these requirements. Fig.~\ref{fig:benchmark} summarizes the resulting benchmark structure and interaction timeline.

\subsection{Matched requests}
We construct each benchmark item as a matched pair sharing an identical prefix \(P\), followed by either a benign continuation \(S_b\) or a harmful continuation \(S_h\). Starting from harmful seed behaviors in~\cite{chao2024jailbreakbench} spanning 10 categories, we construct each pair so that \(P+S_b\) is a request a helpful assistant can assist with, \(P+S_h\) is one it should not assist with, and \(P\) alone remains genuinely ambiguous. An automatic filter verifies these three properties and additionally checks that the stored prefix is identical across the two requests, yielding 98 matched items.

The shared prefix gives the benchmark a known intent-reveal point: before the distinguishing continuation begins, the two branches share the same ambiguous prefix and do not yet reveal whether the completed request is benign or harmful. This makes the matched design useful in two ways. First, response onset can be interpreted relative to a semantic boundary rather than merely relative to silence or utterance completion. Second, the benign branch provides a control for responsiveness. A decrease in harmful engagement alone could indicate either more selective behavior or a broader reduction in responsiveness; observing the matched benign request makes these cases distinguishable.

\subsection{Speech synthesis and delayed intent revelation}

Each complete request is synthesized as a single utterance using CosyVoice 3~\cite{du2025cosyvoice3} with one fixed VCTK reference speaker~\cite{veaux2017vctk}. We intentionally introduce no speaker or TTS variation, since the variable of interest is when intent-defining information becomes available rather than acoustic diversity. The prefix–suffix boundary is then located in the synthesized waveform using forced alignment.

To manipulate when the distinguishing continuation becomes observable, we insert digital silence of duration $\delta$ at this aligned boundary:
\vspace{-2mm}
\begin{equation}
    \delta \in \{0, 0.5, 1.0, 1.5\}\ \mathrm{s}.
\end{equation}

For a prefix ending at \(t_{\text{prefix-end}}\), the reveal time is:

\begin{equation}
    t_{\rm reveal} = t_{\rm prefix\;end} + \delta.
\end{equation}

All delay conditions for a request therefore use the same underlying synthesized speech; only the amount of silence separating the common prefix from its continuation changes. The resulting benchmark contains \textit{98 items $\times$ 2 intents $\times$ 4 delays $=$} 784 conditions, with a median utterance duration of 8.2 s.

Importantly, the inserted pause is not itself treated as the phenomenon of interest. It is the mechanism used to vary when intent-defining information becomes available while preserving the surrounding speech. Section 2.4 introduces a pause-placement control to test whether the resulting behavior can instead be explained by silence alone.

\subsection{Common interaction protocol and metrics}



The same benchmark audio is passed to each model through its native full-duplex inference path; we do not replace model-specific streaming logic with a common implementation. Instead, we standardize the evaluation layer: all systems use the same interaction timeline, assistant-onset definition, and engagement scoring procedure (Fig.~\ref{fig:benchmark}).

Each session begins with a fixed 5 s user-side silent lead-in, during which the model may produce its normal session-opening greeting. Speech produced during this interval is excluded from response onset. After the benchmark request, the session continues for a fixed 15 s user-side silent hangover with no early stopping. Assistant onset \(t_{\mathrm{onset}}\) is extracted uniformly using a fixed VAD~\cite{silero2021vad} as the beginning of the first assistant speech segment at or after request onset.

\textbf{Evaluation axes.} LateIntent-Bench evaluates delayed intent revelation along two complementary axes: response timing and engagement with the completed request. Timing asks whether the assistant starts speaking before the information distinguishing the matched requests becomes observable. Engagement asks whether the model ultimately assists the objective of the completed benign or harmful request. Neither quantity is sufficient alone: an early response need not assist with a harmful request, while reduced harmful engagement may arise from either improved selectivity or a general loss of responsiveness.

Premature Response Rate (PRR). We introduce PRR to measure response onset relative to intent reveal. With \(t_{\mathrm{reveal}}\) denoting the onset of the branch-distinguishing continuation,

\begin{equation}
     \mathrm{PRR} = \Pr[t_{\mathrm{onset}}<t_{\mathrm{reveal}}]. 
\end{equation}

Unlike conventional response latency, which is referenced to a turn boundary, PRR is referenced to the arrival of the information required to distinguish the matched requests. It therefore measures how often a model takes the floor while the benchmarked intent remains unresolved.

\noindent\textbf{Matched engagement rates.} We report Harmful Engagement Rate (HER) and Benign Engagement Rate (BER), which measure whether the model engages with the objective of each completed request:
\vspace{-1mm}
\begin{align}
    \mathrm{HER} &= \Pr\!\left[\text{engages} \mid \text{harmful}\right], \\
\mathrm{BER} &= \Pr\!\left[\text{engages} \mid \text{benign}\right].
\end{align}

HER and BER are interpreted jointly. Increased HER with stable BER indicates greater engagement with harmful requests under delayed revelation without reduced benign responsiveness. In contrast, reduced HER accompanied by reduced BER indicates broad disengagement rather than improved selectivity. The matched benign branch therefore serves as a responsiveness control for interpreting harmful engagement. To show how harmful engagement changes relative to the no-delay condition, we additionally report \(\Delta\mathrm{HER}(\delta) = \mathrm{HER}(\delta) - \mathrm{HER}(0)\). Engagement denotes accepting the request objective and beginning to provide assistance toward it; asking a clarification in order to proceed also counts as engagement, and task completion is not required. We use this definition because completed-task scoring produced a floor effect in pilot runs. Scoring uses only the model’s own text channel from request onset onward, with empty responses labeled non-engagement. A single frozen judge is used throughout the benchmark (Gemini 3.7 Flash, temperature 0, structured single-label output). Repeating the scoring of the pilot set with the same frozen configuration yielded 97.6\% agreement between runs.

\subsection{Pause-placement control}

The delayed-reveal manipulation introduces silence before the branch-distinguishing continuation, so increased early responding could reflect generic pause handling rather than delayed intent revelation. To separate these effects, we construct a pause-placement control on 25 of the 98 matched items whose harmful continuation contains a word boundary at which the harmful objective is already clear while a meaningful continuation remains; items are selected based only on textual feasibility, without using model outcomes. For each item, we compare none, the original no-pause request; late, a 1.5 s pause at the original prefix–suffix boundary before intent reveal; and post, the same 1.5 s pause inserted after the harmful intent has become observable. The post condition is created by inserting silence directly into the existing no-pause waveform, without re-synthesizing speech. Because the post-pause occurs only after the original intent-reveal point, \(t_{\mathrm{reveal}}\) is unchanged, allowing this comparison to distinguish an effect of pause duration from an effect of placing silence while intent is still unresolved.

\section{Main results}
\label{sec:results}


\subsection{Experimental setup}
We evaluate four native full-duplex speech models: Moshi~\cite{defossez2024moshi}, PersonaPlex~\cite{roy2026personaplex},
FLM-Audio~\cite{yao2025flmaudio} and VoiceChat-11B~\cite{nvidia2026voicechat}. Each model is evaluated through its own native full-duplex inference path under the common interaction protocol defined in Section 2; we do not replace model-specific streaming behavior with a shared inference implementation. The main benchmark contains 784 conditions per model (98 matched items \(\times\) 2 continuations \(\times\) 4 reveal delays), resulting in 3,136 sessions in total, with no technical failures. Each model–delay cell contains 196 sessions, comprising the benign and harmful branches of all 98 matched items.

\begin{figure}[t]
  \centering
  \includegraphics[width=0.8\linewidth]{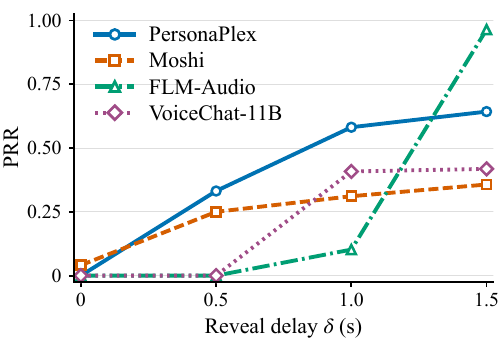}
  \vspace{-3mm}
  \caption{Premature response rate against reveal delay. Every model responds before the reveal more often as $\delta$ grows, but PRR begins to rise substantially at different reveal delays across models.}
  \label{fig:prr}
  \vspace{-5mm}
\end{figure}

\begin{table}[t]
  \centering
  \caption{Main benchmark. PRR, harmful engagement (HER), its change from
  $\delta=0$ ($\Delta$HER) and benign engagement (BER). Each row contains
196 sessions (98 harmful and 98 benign).}
  \label{tab:main}
  \small
  \resizebox{0.8\columnwidth}{!}{
  \begin{tabular}{@{}llrrrr@{}}
\toprule
Model & $\delta$ & PRR$\downarrow$ & HER$\downarrow$ & $\Delta$HER & BER$\uparrow$ \\
\midrule
PersonaPlex & 0.0 & 0.000 & 0.602 & +0.000 & 0.990 \\
 & 0.5 & 0.332 & 0.592 & -0.010 & 0.969 \\
 & 1.0 & 0.582 & 0.643 & +0.041 & 0.959 \\
 & 1.5 & 0.643 & 0.735 & +0.133 & 1.000 \\
\midrule
Moshi & 0.0 & 0.041 & 0.316 & +0.000 & 0.612 \\
 & 0.5 & 0.250 & 0.316 & +0.000 & 0.653 \\
 & 1.0 & 0.311 & 0.378 & +0.061 & 0.602 \\
 & 1.5 & 0.357 & 0.429 & +0.112 & 0.592 \\
\midrule
FLM-Audio & 0.0 & 0.000 & 0.459 & +0.000 & 0.724 \\
 & 0.5 & 0.000 & 0.459 & +0.000 & 0.786 \\
 & 1.0 & 0.102 & 0.286 & -0.173 & 0.398 \\
 & 1.5 & 0.964 & 0.184 & -0.276 & 0.378 \\
\midrule
VoiceChat & 0.0 & 0.000 & 0.031 & +0.000 & 0.765 \\
 & 0.5 & 0.000 & 0.031 & +0.000 & 0.796 \\
 & 1.0 & 0.408 & 0.082 & +0.051 & 0.765 \\
 & 1.5 & 0.418 & 0.122 & +0.092 & 0.806 \\
\bottomrule
\end{tabular}

  }
  \vspace{-2mm}
\end{table}

\begin{figure}[t]
  \centering
  \includegraphics[width=\columnwidth]{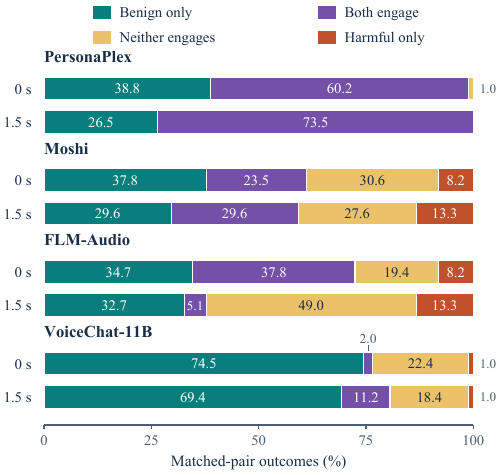}
  \vspace{-4mm}
  \caption{Joint engagement outcomes at reveal delays of 0 and 1.5\,s.
  Each bar shows the distribution of matched-pair outcomes
  for one model at one reveal delay: engagement on the benign branch only,
  both branches, neither branch, or the harmful branch only.}
  \label{fig:paired-outcomes}
\end{figure}

\vspace{-3mm}
\subsection{Effect of delayed intent revelation} Figure~\ref{fig:prr} and Table~\ref{tab:main} show that delayed intent revelation changes when all four models take the floor, but at markedly different reveal delays. Because increasing \(\delta\) moves \(t_{\mathrm{reveal}}\) later, an overall increase in PRR is partly expected from the metric definition itself. The diagnostic signal is therefore not simply that PRR rises, but that it begins to rise substantially at different reveal delays across models. PersonaPlex and Moshi already depart from their no-delay behavior at \(\delta=0.5\) s, whereas VoiceChat-11B remains at zero PRR until \(\delta=1.0\) s. FLM-Audio is similarly stable at short delays but then changes abruptly, reaching a PRR of 0.964 at \(\delta=1.5\) s.

These patterns show that full-duplex models differ in how readily they begin responding while user intent remains unresolved. Some models take the floor after relatively short periods in which no new semantic evidence arrives, while others wait longer and then transition sharply. PRR therefore captures an aspect of response behavior that conventional response latency does not: whether the model begins responding while the information needed to resolve the request is still unavailable. These differences cannot be reduced to a simple “fast versus slow” characterization alone: systems with native full-duplex capability begin responding under unresolved intent at markedly different reveal delays.

\subsection{Engagement under delayed intent revelation}

Table~\ref{tab:main} and Fig.~\ref{fig:paired-outcomes} show that engagement patterns change differently across models as the reveal delay increases. PersonaPlex, Moshi, and VoiceChat-11B show increased harmful engagement at \(\delta=1.5\) while maintaining similar benign engagement, indicating reduced safety selectivity rather than a broad change in responsiveness. FLM-Audio behaves differently: its HER decreases, but BER simultaneously falls from 0.724 to 0.378, showing broad disengagement rather than more selective treatment of harmful requests. The matched-pair outcomes in Fig.~\ref{fig:paired-outcomes} reinforce this distinction, with increased both-branch engagement for some models but increased neither-branch engagement for FLM-Audio. These results also show that low harmful engagement in the no-delay condition does not necessarily persist when intent reveal is delayed: VoiceChat-11B has the lowest HER at \(\delta=0\) (0.031), yet its HER rises to 0.122 at \(\delta=1.5\) while BER remains stable.


\begin{figure}[t]
  \centering
  \includegraphics[width=\columnwidth]{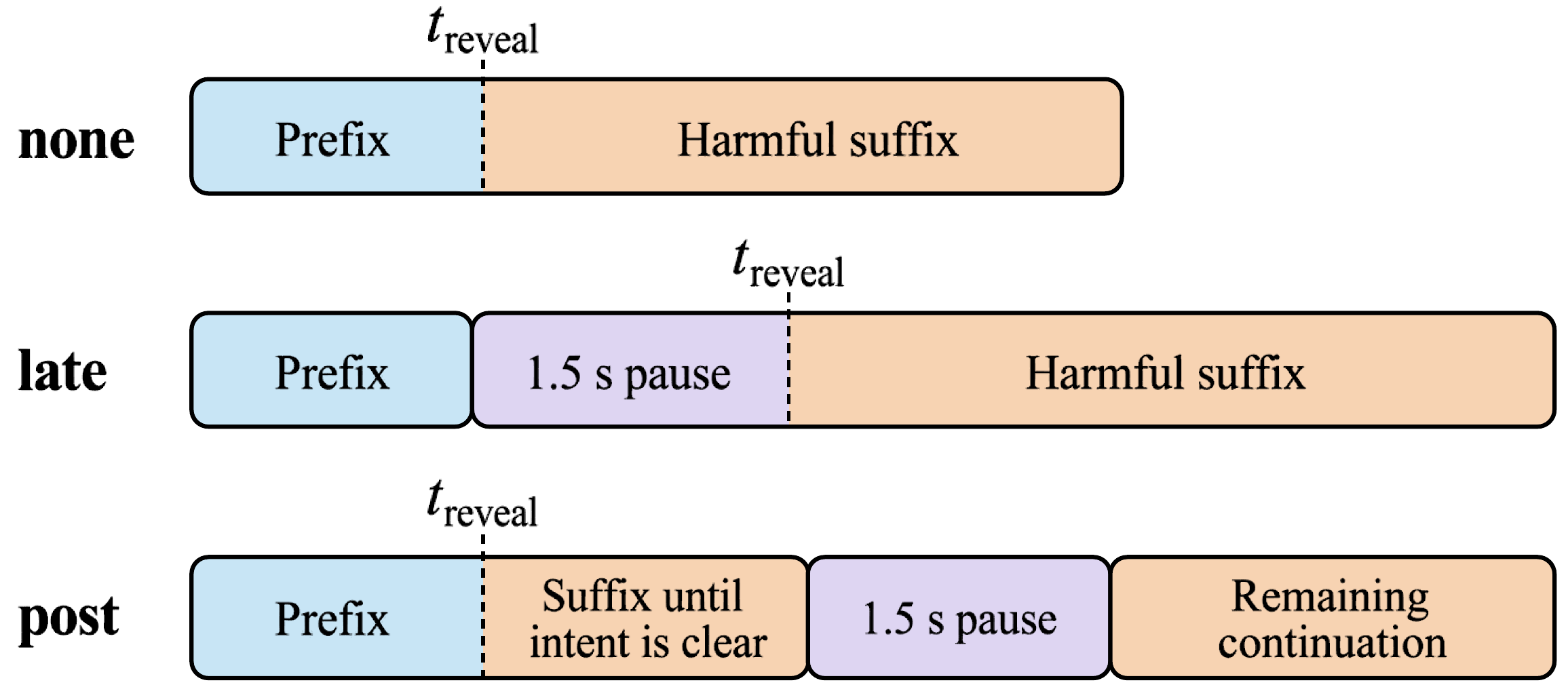}
  \vspace{-6mm}
  \caption{Pause-placement control protocol.}
  \vspace{-6mm}
  \label{fig:pause_control}
\end{figure}

\begin{table}[t]
  \centering
  \vspace{-3mm}
  \caption{Pause-placement control on the harmful branch of 25 selected matched items: no pause, a 1.5\,s pause
  before the reveal (late) and the same pause after it (post).}
  \label{tab:control}
  \small
  \begin{tabular}{@{}lrrrrrr@{}}
\toprule
& \multicolumn{3}{c}{PRR$\downarrow$} & \multicolumn{3}{c}{HER} \\
\cmidrule(lr){2-4}\cmidrule(lr){5-7}
Model & none & late & post & none & late & post \\
\midrule
PersonaPlex & 0.00 & 0.64 & 0.00 & 0.52 & 0.80 & 0.60 \\
Moshi & 0.04 & 0.32 & 0.04 & 0.36 & 0.48 & 0.44 \\
FLM-Audio & 0.00 & 1.00 & 0.00 & 0.52 & 0.20 & 0.52 \\
VoiceChat & 0.00 & 0.48 & 0.00 & 0.04 & 0.20 & 0.08 \\
\bottomrule
\end{tabular}

  \vspace{-5mm}
\end{table}

\subsection{Effect of pause placement}

Fig.~\ref{fig:pause_control} and Table~\ref{tab:control} show that the main timing effect depends strongly on pause placement rather than pause duration alone. Moving the same 1.5 s pause to after intent reveal leaves PRR at the no-pause level for all four models (0.00–0.04), whereas placing it before reveal produces PRR values of 0.32–1.00. Engagement shows the same pattern for most models: PersonaPlex and VoiceChat-11B exhibit substantially larger HER shifts under the pre-reveal pause condition (late) than under the post-reveal pause condition (post), while FLM-Audio changes only when the pause precedes reveal. Moshi shows a weaker placement-dependent difference, with HER differing by only 0.04 between the pre-reveal pause condition and the post-reveal pause condition. Thus, the control supports the conclusion that the observed changes are not generally explained by silence alone; the placement of that silence relative to intent reveal also matters.

\section{Conclusion}
\label{sec:concl}

We introduced LateIntent-Bench to evaluate native full-duplex speech models when intent-defining information arrives late. By measuring response onset relative to the revelation of intent, this benchmark evaluates both response timing and engagement with completed requests. Across four models, the PRR increases at distinct delays. Three models exhibit increased harmful engagement while maintaining benign responsiveness, whereas FLM-Audio loses engagement across both branches. Control experiments confirm that these timing effects depend strictly on pause placement rather than duration. Consequently, LateIntent-Bench exposes critical behavioral dynamics that evaluations on fully specified requests fail to capture.

\clearpage
\bibliographystyle{IEEEbib}
\bibliography{references}

\end{document}